\documentclass[runningheads]{svmult}

\usepackage{makeidx}   
\usepackage{graphicx}  
\usepackage{subeqnar}  
\usepackage{multicol}  
\usepackage{physmubb}  
\makeindex             

\usepackage{epsfig}
\newcommand{\agt}{\,\rlap{\lower 3.5 pt \hbox{$\mathchar \sim$}} \raise 1pt
 \hbox {$>$}\,}
\newcommand{\alt}{\,\rlap{\lower 3.5 pt \hbox{$\mathchar \sim$}} \raise 1pt
 \hbox {$<$}\,}

\begin{document}

\title*{Perspectives on the Standard Model%
\thanks{To appear in the {\it Proceedings of the 15th Topical Conference on
Hadron Collider Physics (HCP 2004)}, 
14 - 18 July, 2004, Michigan State University, East Lansing, Michigan.}
}

\toctitle{Perspectives on the Standard Model}

\titlerunning{Perspectives on the Standard Model}

\author{Sally Dawson\\
Physics Department, Brookhaven National Laboratory, Upton, N.Y.~11973}

\authorrunning{Sally Dawson}

\maketitle              

\begin{abstract}
We discuss recent results from global electroweak fits and
from the Tevatron and review  the motivation for physics at the $TeV$
energy scale.
\end{abstract}

\section{Introduction}
We live in exciting times for particle physics.  The Tevatron 
Run II program is producing  new physics results, which will be 
followed by the startup of the LHC in three years.  
These machines have a rich physics menu of new physics searches,
 precision measurements, QCD
studies, $t$, $b$, and $c$ quark physics, and much more.   

We are in an enviable position.  We have an electroweak theory
which is consistent with all data, but which predicts new physics
waiting to be discovered at the Tevatron and the LHC.  
When the LHC begins accumulating data, the physics landscape will
change dramatically:  the number of observed
top quark pairs will jump from 
$10^4$ at the Tevatron to $10^7$ in the first $10~fb^{-1}$ at the LHC.
Similarly, the number of $b {\overline b}$ pairs in our 
world data set  will increase from
$10^9$ from the $B$ factories  at SLAC and KEK 
to $10^{12}-10^{13}$ at the LHC.
We are confident in our predictions that the wealth of new data at
the LHC will guarantee new discoveries.  
 
In this note, we begin by
reviewing new measurements of the top quark and $W$ boson masses and 
their implications for electroweak physics.  We continue by discussing
the emergence of the Tevatron as a machine for precsion studies.
Finally, we conclude by emphasizing our certainty that new physics
discoveries are just around the corner.

\section{Top and $W$ Masses from the Tevatron}

Two of the most interesting results of the last year were the 
new values for the top quark and the $W$ boson masses. A review of other
electroweak measurements at the Tevatron can be found in
Ref. \cite{topew}.

In 2003, we had for the world average for
the top quark mass\cite{lepew,newt},
\begin{equation}
 M_t =  174 \pm 5.1~GeV \qquad \qquad \qquad \hbox{\bf{2003 Result}}~.
\end{equation}
The D0 collaboration improved their Run I analysis to include matrix
element calculations, and this analysis dominates the new world average.
In 2004, we have the new combination of Run I results\cite{newt},
\begin{equation}
M_t=178.0\pm 4.3 ~GeV\qquad \qquad \hbox{\bf{2004 Result}}
\label{topeq}
\end{equation}
This value for $M_t$ is most sensitive to the D0 lepton plus jets result.
As the Tevatron Run II program  continues, we expect ever more
accurate
measurements of the top quark mass\cite{topup}.

The top quark mass plays 
a special role in the Standard Model.  In QED, the running of $\alpha_{EM}$
at a scale $\mu$ is not affected by heavy quarks with $M>>\mu$.  
The decoupling
theorem tells us that diagrams with heavy virtual particles don't contribute
to experimental observables
at scales $\mu << M$ if two conditions are met:
\begin{itemize}
\item Coupling constants don't grow with $M$, and
\item The gauge theory with the heavy quark removed is still renormalizable.
\end{itemize}
The spontaneously broken $SU(2)\times U(1)$ gauge  theory violates both 
conditions and we expect large effects from virtual diagrams involving the top
quark, with effects from virtual top quarks growing quadratically with $M_t^2$.
A small change in the top quark mass therefore has large effects in
the inferred predictions of the Standard Model. 

The $W$ boson is produced at the Tevatron through the partonic
sub-process, $u {\overline d}\rightarrow W^+$, and the leptonic
decays of the $W$ can be used to determine the $W$ mass and
width to good precision through a measurement of the transverse $W$
mass or the $p_T$ of the lepton.  The Run I value for the $W$ mass
from a combination of CDF and D0 data
 is $M_W=80.452\pm 0.59~GeV$\cite{tevw}.  At LEP2, the $W$ boson
was pair produced through the process, $e^+e^- \rightarrow W^+W^-$,
(see Fig. \ref{eeww_fig}), and a value for the $W$ mass was found of $M_W
=80.412\pm0.042~GeV$.  The Run I and LEP2 values have been 
combined to give a new world average for the $W$ mass of\cite{lepew}
\begin{equation}
M_W=80.425 \pm 0.034 ~GeV\qquad \qquad \hbox{\bf{2004 Result}}~.
\nonumber\\ 
\end{equation}
This is slightly lower than  the previous value,
\begin{equation}
 M_W = 80.426\pm .034~GeV~~\qquad \qquad {\hbox{\bf{2003 Result}}}~.
\end{equation}

\section{Electroweak Precision Measurements}
\subsection{Global fits}

\begin{figure}
\begin{center}
\rotatebox{90}{
\includegraphics[angle=-90,width=.6\textwidth]{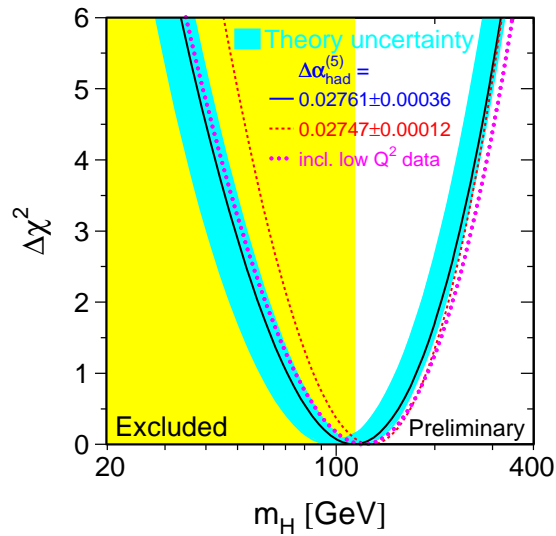}}
\end{center}
\caption[]{ Limit on the Higgs boson mass from the
summer, 2004 LEPEWWG fit, using the updated value for the
top quark mass, Eq. \ref{topeq}\cite{lepew}.}
\label{fg:higsfit}
\end{figure}

The result of the new top quark mass
is to shift the best fit value for the Higgs boson mass as
can be seen in Fig. \ref{fg:higsfit}\cite{lepew,alt},
\begin{eqnarray}
M_h & < &219~GeV \qquad \qquad  \hbox{\bf{2003 Result}}
\nonumber \\
M_h&<&251~GeV \qquad \qquad \hbox{\bf{2004 Result}}
\end{eqnarray}
The small shift in the top quark mass, $\delta M_t \sim 
3~GeV$  shifted the $95\%$ confidence level limit  on the
Higgs mass by roughly
$30~GeV$.  The best fit for the Higgs mass is 
now\footnote{As emphasized in Ref.\cite{alt}, the best fit
value for the Higgs mass depends on whether the measurements
of $M_W$ and $M_t$ from the Tevatron are included in the fit.
A fit to the Higgs mass without the $W$ mass measurement gives
a best fit value of $M_h=129^{+76}_{-50}~GeV$, while a fit
which does not include $M_t$ or $M_W$ as measured at the Tevtraon gives
$M_h=117^{+162}_{-62}~GeV$.}\cite{lepew}
\begin{equation}
M_h=114^{+69}_{-45}~GeV \qquad \qquad{\hbox{\bf{2004 Result}}}~.
\end{equation}
It is reassuring to see that the best fit value is not in the
region excluded by the direct search experiments at LEP. (The 
best fit value of 2003 was $M_h=96^{+60}_{-38}~GeV$, a value
which is excluded by the direct search experiments.) 

\begin{figure}
\begin{center}
\rotatebox{90}{
\includegraphics[angle=-90,width=.6\textwidth]{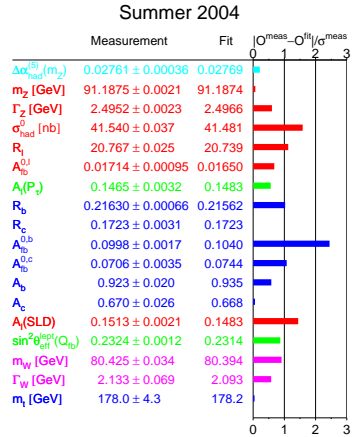}}
\end{center}
\caption[]{Global fit to electroweak data from the 
summer, 2004 LEPEWWG fit.  Low $Q^2$ data is
not included in the fit\cite{lepew}.}
\label{fg:globfit}
\end{figure}

The global fit of the experimental data
to the parameters of the Standard Model is shown in Fig. \ref{fg:globfit}
 and is in spectacular
agreement with the predictions of the Standard Model.
 The fits have been redone using the new 
value of $M_t$
and only the high $Q^2$ data is included in the 2004 fits.  The results
of the low energy experiments such as atomic parity violation in Cesium,
neutrino-nucleon scattering (NuTeV), and Moller scattering ($e^-e^-$) are 
now predicted results. 

The prediction of the global fit for $M_t(fit)=178.2~GeV$ is in good
agreement with the value of Eq. \ref{topeq},
while
 the best fit value for $M_W$ (with the
experimental value of $M_t$ included in the fit) is slightly lower
than the experimental value\cite{alt},
\begin{equation}
M_W=80.386\pm .023~GeV ~~~\qquad \qquad \qquad {\hbox{\bf{Fit}}}~~~.
\end{equation}

\begin{figure}
\begin{center}
\rotatebox{90}{
\includegraphics[angle=-90,width=.6\textwidth]{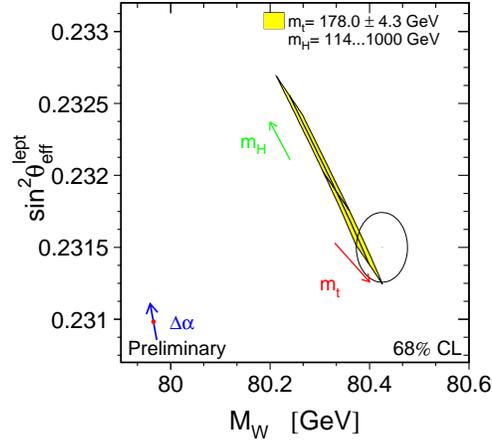}}
\end{center}
\caption[]{Relation between the extracted 
 value of $sin^2\theta^{lept}_{W~eff}$
and the $W$ mass and its dependence on the Higgs boson mass\cite{lepew}.}
\label{fg:swfit}
\end{figure}

The puzzles of the global fit from previous years remain.  The extracted
value of $\sin^2\theta^{lept}_{W~eff}$ still has the problem that measurements
 from  experiments with leptons and from 
experiments with hadrons disagree by $2.9\sigma$\cite{lepew}.  
Using the value of 
$\sin^2\theta^{lept}_{W~eff}$ found from leptons at SLD ($A_l$),
\begin{equation}
\sin^2\theta^{lept}_{W~eff}= 
0.23098\pm 0.00026 
\qquad \qquad {\hbox{\bf{SLD}}},
\end{equation}
 a value of the Higgs boson
mass in conflict with the direct search experiments is predicted.  On
the other hand, using $\sin^2\theta^{lept}_{W~eff}$ from the forward backward
$b$ asymmetry at LEP ($A_{fb}^{0,b}$),
\begin{equation}
\sin^2\theta^{lept}_{W~eff}=0.23212\pm 0.00029 \qquad\qquad
{\hbox{\bf{LEP}}},
\end{equation}
one obtains a prediction in disagreement with 
the measured value of $M_W$.  This can be seen in Fig. \ref{fg:swfit}.
It is extremely 
difficult to construct models which explain this feature.

\begin{figure}
\begin{center}
\rotatebox{0}{
\includegraphics[angle=0,width=.6\textwidth]{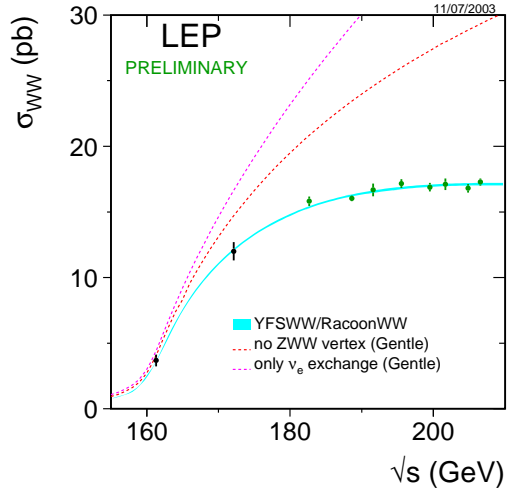}}
\end{center}
\caption[]{The total cross section for the process $e^+e^-
\rightarrow W^+W^-$ measured at  LEP2 and the Standard Model prediction\cite{eeww}.
If the three gauge boson ($ZW^+W^-$) vertex is removed from the
theory, the prediction deviates wildly from the data.}
\label{eeww_fig}\end{figure}

\subsection{Low Energy Measurements}

\begin{figure}
\begin{center}
\rotatebox{-90}{
\includegraphics[angle=0,width=.6\textwidth]{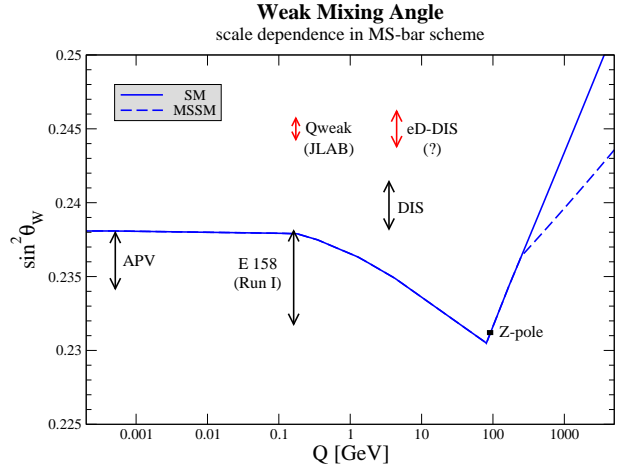}}
\end{center}
\caption[]{ Comparison of low $Q^2$ data from NuTeV, Moller Scattering,
and atomic parity violation with the energy scaling predicted by
the $SU(2)\times U(1)$ Standard Model\cite{erm}. The red points are
proposed future experiments.}
\label{ee}
\end{figure}

Measurements at low energy test our understanding of the energy scaling 
of the theory.  The best fit values for the parameters
of the Standard Model
which are given in Fig. \ref{fg:globfit} can be used to make predictions for
Moller scattering, for the value of
$\sin^2\theta_W$ which is extracted from NuTeV experiment, 
and for atomic parity violation in Cesium, and then compared
with the experimental values. 
The energy dependence of the weak mixing angle $\sin^2\theta_W$ is
predicted in the Standard Model
and  is shown as the solid line in 
Fig. \ref{ee} and we see that the 
results of the 
low energy experiments
are in reasonable, although not perfect,agreement with the predictions
of the Standard Model except for the NuTev result\cite{erm}.  

The  NuTeV  collaboration has
published its result for the ratio
of neutral currents to charged currents
in neutrino-nucleon scattering.  When
the measurement is interpreted 
in terms of  $\sin^2\theta_W$, it
is $3\sigma$ away from the global fit\cite{lepew}:
\begin{eqnarray}
\sin^2\theta_W&=& 0.2277\pm 0.0013(stat)\pm 0.0009(syst)
-0.00022{M_t^2-(175~GeV)^2\over (50~GeV)^2}
\nonumber \\ && +0.00032\ln\biggl(
{M_h\over 150~GeV}\biggr)~~~~~~{\hbox{\bf{Experiment}}}\nonumber \\
\sin^2\theta_W&=&0.2227\pm 0.0004\qquad\qquad \qquad {\hbox{\bf{Fit}}}~~~.
\end{eqnarray}
The NuTev experiment is under intense theoretical scrutiny since connecting
the experimental observables with theoretical predictions requires a
detailed understanding of theory. 
A number of possible solutions to the discrepancy 
between theory and experiment have been proposed,
including an asymmetry in the strange/anti-strange parton 
distributions\cite{kretzer},
and the effects of higher order ${\cal O}(\alpha)$ corrections\cite{diener}

The situation with atomic parity violation has improved since the 2002
global fit showed a $1.5\sigma$ deviation of the experiment from the global
fit.  There is a new calculation of QED corrections which is now in
good agreement with the best fit value\cite{kuchiev}
\begin{eqnarray}Q_W&=-72.84\pm 0.29 (exp) \pm 0.36 (theory)
\qquad&{\hbox{\bf{Experiment}}}
\nonumber \\
Q_W&= -72.880\pm 0.003\hskip .55in \qquad \qquad \qquad &{\hbox{\bf{Fit}}}~~.
\end{eqnarray}

Moller scattering ($e^-e^-\rightarrow e^-e^-$) provides a clean measurement
of the weak mixing angle since it is a purely leptonic process.
A cominbination of the E158 Experiment at SLAC Run
I,II and III data gives the result\cite{e158},
\begin{eqnarray}
&&\sin^2\theta_W(Q^2=.026~GeV^2)=\nonumber \\
&& \quad 0.2403\pm 0.0010(stat)\pm 0.0009(stat)\qquad~~~~~~
 {\hbox{\bf{Experiment}}}
\end{eqnarray}
In good agreement with the theoretical prediction,
\begin{equation}
\sin^2\theta_W(Q^2=.026~GeV^2)=
0.2385 \pm 0.006\qquad\qquad {\hbox{\bf{Theory}}}~~.
\end{equation}

The inclusion of the low $Q^2$ results into the fit has 
very little effect on the Higgs mass limits.
The conclusion  from the electroweak fits is that
electroweak physics is in even better shape this year than last year!

\section{The Standard Model at the Tevatron}
The Standard Model continues to be tested at the Tevatron. In
the coming few years, we can expect that
the properties  of Standard Model particles will be
measured with high precision.  $W$ and $Z$ gauge
bosons have large cross sections at the Tevatron, so high statistics and precision
measurements will dominate:  $W$ and $Z$ masses and  width
measurements, production cross sections,
and gauge boson pair production cross sections.
Measurements of the top quark mass and top quark properties will continue to unfold,
along with
Higgs searches, new physics searches, $b$ physics, and QCD studies, to name just
a few areas. This conference saw a large number of new results
in these areas  from the D0 and
CDF collaborations.

The top quark plays a leading role in many models  with physics
beyond the Standard Model, so measurements of top quark properties
play an especially important role in constraining these models. 
One particularly important measurement which we look forward to in the 
future 
is single top production.  Single top production
at the Tevatron has a total rate
of roughly $\sigma\sim 3~pb$\cite{singletoptheory}, 
approximately  half that of top
pair production, and can serve to measure the CKM mixing
 parameter, $V_{tb}$. D0 has looked for single top
production in $156~pb^{-1}$ and $169~pb^{-1}$ of Run II data and
finds the $95\%~c.l.$ limits\cite{singletopd0},
\begin{eqnarray}
\sigma({\hbox{s-channel~ production}}) & < & 19~pb\nonumber \\
\sigma(\hbox{t-channel~ production})& <& 25 ~pb\nonumber\\
\sigma(\hbox{s+t-~ channel~ production})
 &<& 23~ pb~~\quad {\hbox{\bf{D0}}}~,
\end{eqnarray}
while CDF finds the corresponding $95\%~c.l.$ limits\cite{singletopcdf},
\begin{eqnarray}
\sigma(\hbox{s-channel~ production})&<& 13.6~pb\nonumber \\
\sigma(\hbox{t-channel~ production})& <& 10.1 ~pb\nonumber\\
\sigma(\hbox{s+t-~ channel~ production})&<&~ 17.8~ pb~~\quad {\hbox{\bf{CDF}}}~.
\end{eqnarray}

\begin{figure}
\begin{center}
\rotatebox{0}{
\includegraphics[angle=0,width=.6\textwidth]{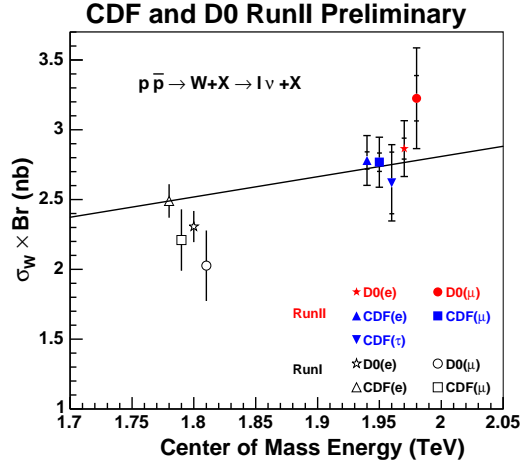}}
\end{center}
\caption[]{Comparison of CDF and D0 results for $W$ production,
followed by the decay $W\rightarrow l\nu$\cite{wzprod_d0}.}
\label{figw}
\end{figure}

\begin{figure}
\begin{center}
\rotatebox{0}{
\includegraphics[angle=0,width=.6\textwidth]{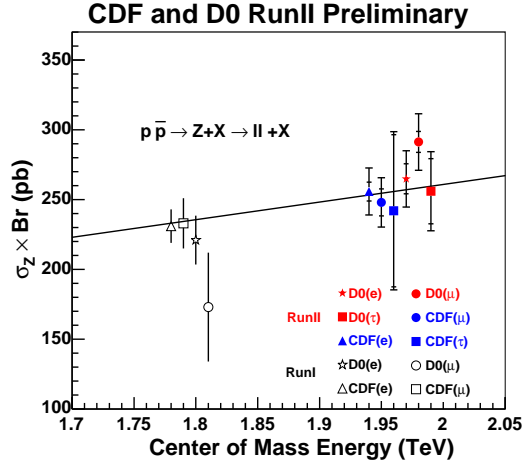}}
\end{center}
\caption[]{Comparison of CDF and D0 results for $Z$ production,
followed by the decay $Z\rightarrow l^+l^-$\cite{wzprod_d0}.}
\label{figz}
\end{figure}

The production of $W$ and $Z$ bosons
 tests the QCD production mechanism, while
the ratio of the production cross section times the leptonic branching
ratios can be used to extract the $W$ boson total width.  Figs. \ref{figw} and \ref{figz}
show the results of $177.3~pb^{-1}$ of data from D0, and $72~pb^{-1}$ from CDF,
 for $W$  production with the decay
$W\rightarrow l \nu$ and for $Z$ production followed by the leptonic decay, $Z\rightarrow
l^+l^-$, along with a comparison to the Standard Model prediction. There
is good agreement with the Standard Model predictions.
 
A particularly interesting set of results concerns two gauge boson production, which
is sensitive to the three gauge boson vertices, and hence to the non-Abelian nature
of the theory. 
The results from LEP2 for $e^+e^-\rightarrow W^+W^-$ 
(Fig. \ref{eeww_fig}) show that the
three gauge boson couplings are very close to their Standard Model values.  
In an effective theory where there is new physics
at a higher energy scale, the non-Standard Model contributions to the
three gauge boson  couplings
would grow like ${s\over \Lambda^2}$\cite{effec} and we can expect
enhanced couplings at the Tevatron and LHC.  Preliminary measurements\cite{wwsig_d0}
 of the $W^+W^-$
pair production cross section at the Tevatron show good agreement with next-to-leading
order (NLO) QCD predictions and no indication
of non-Standard Model gauge couplings\cite{wwnlo},
\begin{eqnarray}
\sigma_{WW}&= 13.8^{+4.3}_{-3.8}(stat)^{+1.2}_{-0.9}(syst)\pm 0.9~pb~~~~&{\hbox{D0}}
\nonumber \\
\sigma_{WW}&= 14.3^{+5.6}_{-4.9}(stat)\pm 1.6 (syst)\pm 0.9~pb~~~~&{\hbox{CDF}}
\nonumber \\
\sigma_{WW}&= 12-13.5~pb~~~~~~&{\hbox{NLO~theory}}\quad .
\end{eqnarray}

\subsection{Advances in theory}
Advances in theory go hand in hand with higher
statistics measurements at
the Tevatron.  The strong coupling constant is not small: 
\begin{equation}
 \alpha_s(M_Z)
\sim .12 >> 15 \alpha_{EM}~.
\end{equation}
  Emission of soft gluons 
and  multi-particle states are important, 
along with
large effects from higher order perturbative QCD
corrections. Understanding precision measurements
at the Tevatron and the LHC
requires new tools  and the systematic inclusion of higher
order QCD effects.  This in turn requires new techniques for
computing the  higher order corrections.

\begin{figure}
\begin{center}
\rotatebox{0}{
\includegraphics[angle=0,width=.6\textwidth]{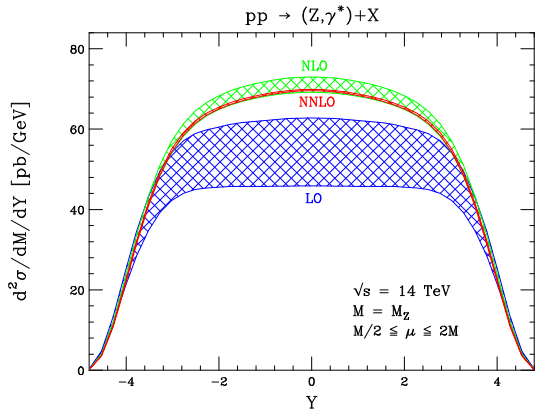}}
\end{center}
\caption[]{Rapidity distribution of on-shell $Z$ bosons at the LHC. 
 The band indicates the
 renormalization/factorization scale dependence\cite{ADMP}.}
\label{drellyan_fig}
\end{figure}

Many processes are now calculated at next-to-next-to leading order
(NNLO).  Not only  total cross sections\cite{nnlo},
but also kinematic distributions  for a few processes
are becoming available at NNLO.
For example, predictions for 
the Drell-Yan rapidity computed 
to NNLO in perturbative QCD at the LHC are shown
in Fig. \ref{drellyan_fig}\cite{ADMP},
We see the very small error bar due to the scale dependence at NNLO.  This suggests that it may be
possible to use Drell Yan production to measure parton distribution
functions at NNLO. 

\begin{figure}
\begin{center}
\rotatebox{0}{
\includegraphics[angle=0,width=.6\textwidth]{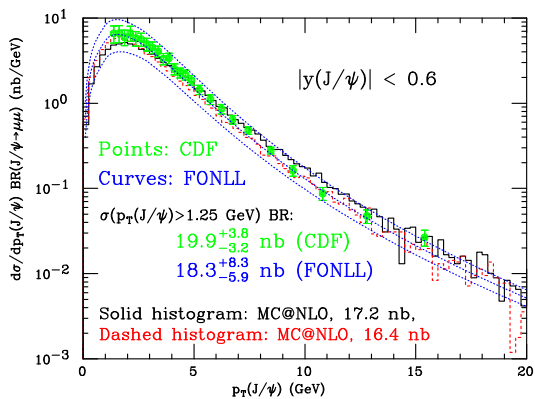}}
\end{center}
\caption[]{$b$ pair production
cross section at the Tevatron.  The theoretical predictions
include higher order effects in both the production and the 
decay and use the Monte Carlo program MC@NLO\cite{bnlo}.}
\label{bfig}
\end{figure}

Understanding the precise data from the Tevatron and future
data from the LHC  will require
Monte Carlo programs which incorporate physics at next-to-leading order
(NLO) in perturbative QCD.
There are several programs on the market.
The MC@NLO program matches parton showering calculations 
at low $p_T$ with exact
NLO matrix element calculations at high $p_T$\cite{mcnlo}. 
An example of the importance of
properly incorportating higher order effects is the $b$ production
cross section at the Tevatron, which has been a long standing puzzle.  By
including the higher order QCD corrections in both the production and the
fragmentation functions, the theoretical prediction now agrees quite well
with the data, as can be seen in Fig. \ref{bfig}\cite{bsigs}.   

Another approach to including higher order QCD corrections is the MCFM
Monte Carlo program\cite{mcfmref}. This program includes exact matrix
element calculations for both the signal and background processes for
a number of processes at hadron colliders.  For most processes, the matrix
elements are included to next-to-leading order and the full set of spin
correlations is included for the decays.

\section{Outlook}
Despite the successes of the Standard Model, the arguments
for new physics at the
$TeV$ scale have never been stronger. Many of the questions particle
physicists seek to answer become even more pressing as accelerators
approach the $TeV$ energy scale:
\begin{itemize}
\item
 The question of  naturalness:   Why is $M_W$ so much
smaller than $M_{pl}$?
\item
Dark matter:  What is it?  Why is there so much of it?
\item Neutrino masses:  Where do they come from?  Why are they so small?
\item What mechanism restores unitarity to the
$WW$ scattering amplitude:  Is there a light
Higgs?  Is there some other mechanism?
\end{itemize}
Extensions of the Standard Model which have been proposed in various
attempts to answer these questions almost uniformly predict physics
beyond that of the Standard Model at the $TeV$ energy 
scale\cite{lh}. So even though current results from LEP and the Tevatron
support the experimental validity of the Standard Model, we are 
confident that there is new physics just around the corner.

Furthermore, the Higgs sector of the Standard Model is unsatisfactory to
theorists.
Light scalar particles, such as the Higgs boson, are unnatural in
the sense that the scalar Higgs mass depends quadratically on new physics
which may exist at
a higher energy scale.  In order to have a Higgs boson mass below $200~GeV$
(as suggested by the precision electroweak measurements),
large cancellations between various contributions  to the
Higgs boson mass are required.  Attempts
to avoid this problem have led to an explosion of model building in
recent years\cite{lh,newmods}.

Models of new physics are highly constrained by precision
measurements.  New physics effects can be parameterized in a model
independent fashion by adding a tower of dimension six operators
to the Standard Model:
\begin{equation}
{\cal L}\sim \Sigma {c_i\over \Lambda^2} {\cal O}_i~~~.
\end{equation}
Precision measurements place significant limits on many possible
operators.  For example,current data requires\cite{guid},
\begin{eqnarray}
{\cal O}_i =
{\overline e}\gamma_\mu e{\overline l}\gamma^\mu l &~~~~\Lambda >&4.5 - 6 ~TeV
\nonumber \\
{\cal O}_i={\overline e}\gamma_\mu \gamma_5 e{\overline b}\gamma^\mu \gamma_5 b
 &~~~~\Lambda >&3-4 ~TeV~.
\label{ops}
\end{eqnarray}
A more complete list of constraints on dimension 6 operators is
given in Ref. \cite{effec,guid}. 
Operators which violate flavor conservation are even more tightly
constrained, often requiring scales $\Lambda > 100~TeV$.
 The fact that physics at the $1~TeV$ scale may
already be excluded experimentally is termed the ``little hierarchy'' problem.

Attempts to evade the little hierarchy probem have recently
stimulated much activity in electroweak scale
model building.  These attempts fall into three general categories:
\begin{itemize}
\item
Remove  the Higgs boson  completely: Models of this
type include models with  dynamical symmetry breaking\cite{simmons} and 
Higgsless models in extra dimensions\cite{higgsless}.
\item
Lower the cut-off scale $\Lambda$ of Eq. \ref{ops}: Examples of this class of
models are models with large extra
dimensions\cite{ld}. 
\item
Force cancellations between contributions to the Higgs
boson mass:  Supersymmetric models are the favorite
of this type, while little Higgs models are a new entrant into
the theory game (although they have difficulty being
consistent with precision measurements\cite{litt}.)
\end{itemize}
Each of these models makes different predictions for physics at
the $TeV$ scale.
Only the data can tell for sure what lies ahead!

\end{document}